\documentclass[aps,prl,twocolumn,groupedaddress,showpacs]{revtex4}
\usepackage{graphicx}
\usepackage{amssymb}
\usepackage{bm}
\begin{document}

\title{Matter-wave gap solitons in atomic bandgap structures}

\author{Elena A. Ostrovskaya$^{1,2}$ and Yuri S. Kivshar$^1$}

\affiliation{$^1$Nonlinear Physics Group, Research School of
Physical Sciences and Engineering, The Australian National University,
Canberra ACT 0200, Australia \\
$^2$Department of Physics and Theoretical Physics, The Australian
National University, Canberra ACT 0200, Australia}

\begin{abstract}
We demonstrate that a Bose-Einstein condensate in an optical
lattice forms a reconfigurable matter-wave structure with a
band-gap spectrum, which resembles {\em a nonlinear photonic
crystal} for light waves. We study in detail the case of a
two-dimensional square optical lattice and show that this {\em
atomic bandgap structure} allows nonlinear localization of atomic
Bloch waves in the form of two-dimensional matter-wave gap solitons.
\end{abstract}

\maketitle

Recent experiments \cite{Arimondo1_02,Esslinger_01,3d_latt_02,Phillips_02} have demonstrated that a Bose-Einstein condensate (BEC) loaded into an optical lattice is a
perfect test ground for a range of fascinating physical effects.
In particular, this is because the effective potential of an
optical lattice can be easily manipulated by changing the
geometry, polarization, phase, or intensity of the laser beams.
Due to inherent nonlinearity of the coherent matter waves
introduced by inter-atomic interactions, BEC in a lattice potential can form a periodic
nonlinear system, which is expected to display rich and complex dynamics.

On the other hand, propagation of light in dielectric structures
with a periodic variation of the refractive index is receiving
growing attention. Photonic bandgap (PBG) materials
\cite{pbg} -- artificial periodic structures with a high index contrast -- can be used to
effectively control the flow of light. The study of photonic
crystals made of a Kerr nonlinear material, the so-called {\em
nonlinear photonic crystals},  has revealed that such structures
exhibit a wealth of nonlinear optical phenomena and, in
particular, they can support self-trapped nonlinear localized modes of the
electromagnetic field in the form of optical gap solitons
\cite{gap_sol_2d,egg}. Dynamically reconfigurable PBG structures - optically-induced refractive index gratings in nonlinear materials - are now offering new ways to control light propagation and localization \cite{moti}.

As demonstrated in this letter, BEC in an optical lattice can be regarded
as a {\em fully reconfigurable} analog of a nonlinear photonic crystal for
matter waves - an "atomic band gap" (ABG) structure. A deep
analogy between coherent light and matter waves suggests that the
concepts employed in the study of nonlinear PBG structures
 can be borrowed for analyzing BEC in optical lattices. Many properties of ABG structures can potentially be exploited for high-precision control and manipulation of coherent matter waves in a similar way to how PBG structures are used to manipulate light. The effect of nonlinear localization in bandgaps is one of these properties.

Localized nonlinear excitations of coherent matter waves - {\em bright
atomic solitons} - could be very useful for applications such as
atomic interferometry due to their robust nature. However, their
generation has so far been experimentally achieved only in BECs with attractive
atomic interactions (analogous to self-focusing optical media)
which are unstable against collapse above a small  ($\sim 10^3$)
critical number of particles.

In theory, a shallow 1D optical lattice can support bright matter-wave solitons
even in repulsive BEC \cite{gap_1d} with large atom numbers. These solitons, described in a framework a coupled-mode theory \cite{gap_1d}, are localized on a large number of lattice wells, and are predicted to exist only in atomic bandgaps.  In the opposite case of a deep lattice, the condensate can be described by the superposition of ground
state modes in the individual wells. The mean-field treatment of the condensate in this regime leads to a discrete equation which admits solutions in the form of stationary
modes localized on a few lattice sites - {\em discrete
solitons} \cite{discrete_1,discrete_2}, in complete
parallel with spatial solitons in periodic optical
structures \cite{discrete_optics} and localized modes
of atomic lattices \cite{discrete_lattice}.

The theory of the nonlinear localized matter waves in optical
lattices is mostly limited to 1D case \footnote{After submission of this manuscript, the generation of   2D arrays of localized states via modulational instability of Bloch waves was theoretically demonstrated in B.B. Baizakov, V.V. Konotop, and M. Salerno, J. Phys. B: At. Mol. Opt. Phys.  {\bf 35}, 5105 (2002)}. However, as can be deducted from the analogous studies of 2D PBG structures \cite{gap_sol_2d}, the nonlinear localization of in BEC in higher-dimensional lattices is qualitatively different because both the symmetry and dimensionality of the lattice start to play an important role in
the formation and properties of the bandgap structure and
corresponding nonlinear modes. In this Letter, we analyze the as yet
unexplored problem of the existence and stability of 2D matter-wave
gap solitons of BEC with repulsive interatomic interactions loaded
into optical lattices, which are analogous to localized states of
light waves in nonlinear photonic crystals. We show that their
accurate description is only possible within a full mean-field
model of BEC in a periodic lattice potential, beyond the
tight-binding approximation or coupled-mode theory often employed
for the study of BEC in 1D optical lattices. 

The dynamics of a three-dimensional BEC cloud  loaded into a
two-dimensional optical lattice can be described by the
Gross-Pitaevskii (GP) equation for the macroscopic wavefunction, $\Psi(\bm{r},t)$ of
the condensate,
\begin{equation}\label{3d}
i\frac{\partial\Psi}{\partial t} =
\left[-\frac{1}{2} \nabla^{2} + V(\bm{r}) +
\gamma_{\rm 3D} |\Psi|^{2}\right]\Psi.
\end{equation}
This equation has been made dimensionless by using the natural
length, time, and energy scales of the optical lattice. The
characteristic length scale in this model, $a_L=k^{-1}_L=d/\pi$,
defines the size of a single well of the lattice potential through
the lattice constant that depends on the
laser wavelength as $d=\lambda/2$. The depth of the
lattice potential, measured in the lattice recoil energy
$E_L=\hbar^2/(2m a^2_L)$, is proportional to the laser intensity
and can be varied from $0$ to $\sim 20 E_L$ in experiment. The
characteristic frequency scale is $\omega_L=E_L/\hbar$. The
interatomic interactions are characterized by the coefficient
$\gamma_{\rm 3D}=8\pi(a_s/a_L)$, where $a_s$ is the $s$-wave
scattering length, positive for
repulsive interactions.
\begin{figure}[htb]
\includegraphics[width=8.5cm]{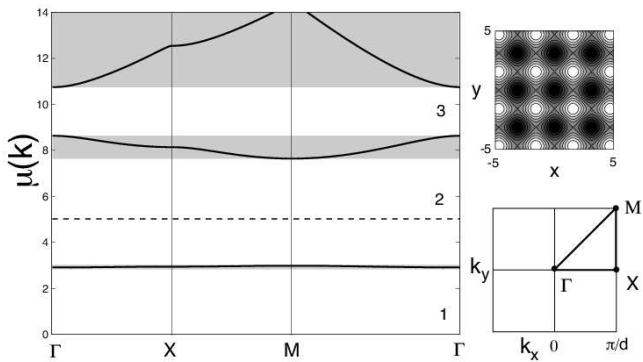}
\caption{\label{dispersion} Dispersion diagram for atomic Bloch
waves in a 2D lattice in the reduced zone representation: shaded -
first three energy bands, numbered - band gaps ($V_0=5.0$). Dashed -
the line $\mu=V_0$. Right top: contour plot of the lattice
potential in Cartesian space, black shading corresponds to
potential minima. Right bottom - the first Brillouin zone of the 2D
lattice in the reciprocal lattice space, marked are the
high-symmetry points of the irreducible zone.}
\end{figure}
\begin{figure}[hbt]
\includegraphics[width=7.5cm]{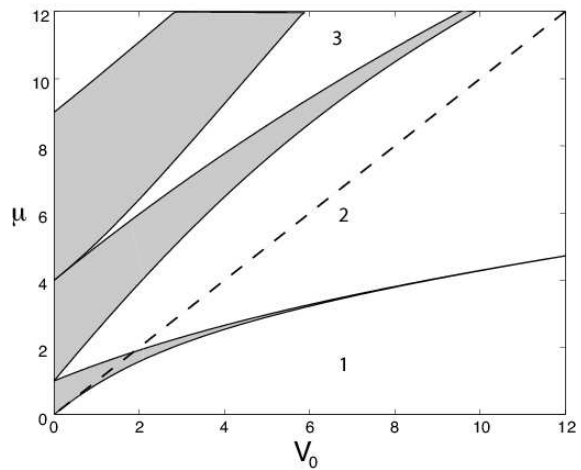}
\caption{\label{bandgaps} Band-gap structure of the atomic
Bloch-wave spectrum as a function of the lattice amplitude $V_0$.
Shaded - Bloch-wave bands, numbered - gaps. Dashed - the line
$\mu=V_0$. }
\end{figure}

In the simplest case of a square optical lattice, the combined
potential of the lattice and magnetic trap, $V(\bm{r})=V_L(\bm{r})+V_T(\bm{r})$, can be written as 
$V(\bm{r}) = V_0 \left(
\sin^{2}x+ \sin^{2}y \right) + { \omega_i}^2\bm{r}^2/2,$
 where $V_0$ is the amplitude of
the optical lattice, and $\omega_i$ are the (normalized) trapping
frequencies in the corresponding directions.  The current
experiments with 2D optical lattices are performed with a strongly
anisotropic BEC cloud, with the ratio of the trapping strengths
$\Omega=\omega_x/\omega_{y,z} \sim 10^{-1}$
\cite{Esslinger_01}. To simplify our analysis, we assume that the
relatively weak magnetic confinement characterized by trap frequencies
$\omega_i$ has small effect on the stationary states of the
condensate in the lattice. Under this assumption, the trap
component of the confining potential in the lattice plane can be
neglected, and the model can be reduced to a two-dimensional GP
equation by assuming separable solutions of the form,
$\psi(\bm{r};t)=\Phi(z)\psi(x,y;t)$ where 
the function $\Phi(z)$, describing the condensate in the direction perpendicular to the
lattice, is a solution of the 1D quantum
harmonic oscillator problem, with the normalization condition
$\int \Phi^2(z) dz =1$. Then the standard dimensionality reduction
procedure \cite{pearl}, leads to the
following equation for the 2D condensate wavefunction in the 2D
lattice potential:
\begin{equation}\label{dimless}
\label{eq2D}
i\frac{\partial\psi}{\partial t} =
\left[-\frac{1}{2}\nabla_{\perp}^{2} + V_{\rm L}(\bm{r}) +
 |\psi|^{2}\right]\psi,
\end{equation}
where $\nabla_{\perp}=\partial^2/\partial x^2+\partial^2/\partial
y^2$, $V_{\rm L}(\bm{r})\equiv V_{\rm L}(x,y)$ is the periodic potential of the
optical lattice, and the wavefunction is rescaled as $\psi \to
\psi \sqrt{\gamma_{\rm 2D}}$, with $\gamma_{\rm 2D}=\gamma_{\rm
3D}/\sqrt{2}$.

Spectrum of atomic Bloch waves in the optical lattice can
be found by exploiting the analogies with the theory of
single-electron states in crystalline solids. Stationary
 states of the condensate in an infinite
periodic potential of a 2D optical lattice are described by
solutions of Eq. (\ref{eq2D}) of the form: $\psi (\bm{r},t) =
\phi(\bm{r}) \exp (-i\mu t)$, where $\mu$ is the chemical
potential, corresponding to the energy level of the stationary
state in the lattice potential. The case of noninteracting
condensate formally corresponds to Eq. (\ref{dimless}) being
linear in $\psi$. According to the Bloch theorem, the stationary
wavefunction  can then be sought in the form
$\phi(\bm{r})=u_{\bm{k}}(\bm{r})\exp(i\bm{k}\bm{r})$, where the
wavevector $\bm{k}$ belongs to a Brillouin zone of the square
lattice, and $u_{\bm k}(\bm{r})=u_{\bm k}(\bm{r}+\bm{d})$
is a periodic (Bloch) function with the periodicity of the
lattice. For the values of $\bm{k}$ within an $n$-th Brillouin
zone, the dispersion relation for the 2D Bloch waves,
$\mu_{n}(\bm{k})$, is found by solving the  following linear
eigenvalue problem
\begin{equation}
\label{eigen}
 \left[ \frac{1}{2}(-i \bm{\nabla}_\perp+\bm{k})^2 +
V_L(\bm{r})\right] u_{n,\bm{k}} = \mu_{n}(\bm{k})
u_{n,\bm{k}}.
\end{equation}
The eigenvalue problem (\ref{eigen}) is simpler than a general 2D
problem due to the separability of the lattice potential. The
typical dispersion relation for the energy of the Bloch states in
the lowest Brillouin zones is shown in Fig. \ref{dispersion}
for a moderate value of the lattice depth $V_0=5$. The
dispersion diagram is presented in the reciprocal lattice space,
and  the dispersion relations are calculated along the
characteristic high-symmetry directions of the irreducible
Brillouin zone (see Fig. \ref{dispersion}, bottom right). It can
be seen that the absolute bands and gaps are determined by the
spectra of the Bloch waves in the middle (point $\Gamma$) and on
the edge (point ${\rm M}$) of the Brillouin zone.

To understand how the variation in the lattice parameters affects the band structure, we note that, owing to the scaling properties of the model, any increase in the well spacing, $a_L$, translates into decreasing the well depth $V_0$. Thus the global behaviour of the band structure can be understood by examining the Bloch wave spectrum as a function of the lattice depth. 
Figure \ref{bandgaps} presents the chemical potentials corresponding to the atomic Bloch waves at the edges of the absolute bands plotted as functions of $V_0$.
The dashed line $\mu=V_0$ (dashed line in
Fig. \ref{dispersion}) separates the quasi-unbound ($\mu>V_0$)
and strongly bound ($\mu < V_0$) condensate states in the
lattice (see also Ref. \cite{Drese}). By varying the interwell separation or the amplitude of the lattice, different regimes of the atomic bandgap structure can be accessed. The tight-binding
regime of the condensate dynamics corresponds to the domain $V_0 \gg V^*_0=\mu$, where the bands "collapse" to discrete levels of bound states in a single isolated well (see Fig. \ref{bandgaps}). The narrow-gap regime of a shallow
lattice, where the coupled-mode theory applies, is found for the
opposite case, $V_0 \ll V^*_0$.
\begin{figure}[h]
\includegraphics[width=7.5 cm]{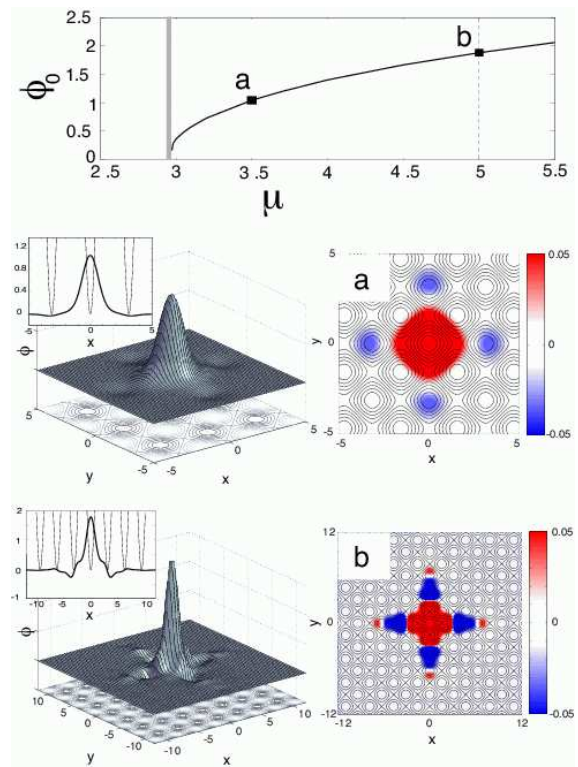}
\caption{\label{modes} Peak amplitude, $\phi_0 \equiv \phi(0,0)$, of the fundamental gap soliton in the first gap ($V_0=5.0$); shaded - the first band, dashed - line $\mu=V_0$. Below: Spatial structure of the gap solitons (a) near the high-energy edge of the first band, $\mu=3.5$, and (b)
at the top of the lattice potential, $\mu=5.0$. Contour plots on the 3D plots (left) show the structure of the lattice potential, darker contours correspond to potential minima. Potential contours on the right are an eyeguide. Insets: the cross-sections of the localized modes and the potential along the line $(x,0)$.  }
\end{figure}

The study of the 1D atomic bandgap structures \cite{Alfimov_02,pearl} indicates that the correct structure and 
dynamical properties of the nonlinear localized modes can only be
revealed through the analysis of the full continuous mean-field GP
model that bridges the gap between the coupled-mode theory and
discrete tight-binding approach. One promising development in
approximate analysis would be generalization of the Wannier function
formalism \cite{wannier_02} to the 2D case. However, the complexity involved  in
the generation of optimally localized higher-dimensional Wannier
functions  \cite{wannier_2d} may not provide significant advantage compared
with the direct solution of the model equation (\ref{dimless}).

Here, we find spatially localized stationary solutions of Eq.
(\ref{dimless}) numerically. Our numerical procedure involves
minimization of the functional $\mathbb{N}=\int f^\dag f {\rm d} \bm{r}$,
where $f(\phi)=[\nabla^2_{\perp}+\mu-V_L(\bm{r})-|\phi|^2]\phi$, by
following a descent technique with Sobolev preconditioning
\cite{jj}.  The minimization procedure yields a stationary state
when $\mathbb{N}(\phi)=0$. Previous applications of this method to
the analysis of optical solitons has shown that it is suitable for
tackling both the fundamental and higher-order nonlinear localized
modes. Our numerical method is independent of the lattice symmetry; it offers a versatile technique for finding and analyzing nonlinear localized modes in a range of different bandgap structures,  and for making accurate predictions about their density, degree of localization, and spatial structure in different areas of the parameter space.

We have identified different families of matter-wave gap solitons
of the repulsive BEC in the 2D atomic bandgap structure.
These solitons can exist in all gaps, excluding the
semi-infinite gap of the spectrum (marked by
1 in Figs \ref{dispersion}, \ref{bandgaps}) below the first band. We note in passing that only conventional nonlinear localization of BEC with attractive interactions is possible in the semi-infinite gap. The change in the peak amplitude of the the  lowest-order  "fundamental" gap solitons across the first gap, along with their spatial localization properties, are demonstrated in Fig. \ref{modes}, for $V_0=5$.  Gap solitons of repulsive BEC branch off the lower-energy (i.e. smaller values of $\mu$) gap edge (see Figs. \ref{modes}, top). Near this edge, i.e. at low particle numbers and amplitudes,  their central peak is localized around a single well [Figs. \ref{modes} (a)].  Further within the gap, the localized mode of the BEC contains larger numbers of atoms,  and has a strongly localized central peak. Near the linear tunnelling threshold, $\mu \sim V_0$, the nonlinear localized states develop extended tails in the orthogonal directions $(x,y)$, along which the tunnelling is assisted by the lower inter-well barrier heights (compared to $2V_0$ in diagonal directions). Although the stability analysis
of the states localized on the extended background (for $\mu>V_0$) is difficult
due to the large computational domain, we have confirmed that
strongly localized 2D modes are dynamically stable. In the lattice geometry under consideration, the line $\mu=V_0$ lies within the first gap (marked 2 in Figs. \ref{dispersion}, \ref{bandgaps}), and only localized modes within this gap have been analyzed.

The higher-amplitude gap solitons have a nontrivial
structure of the tails [see Fig. \ref{modes} (b)], i.e. the zeros of the matter wave are centered within individual lattice wells (see Figs. \ref{modes}, insets), rather than on the potential maxima. This structure is determined by the spatial structure of the Bloch wave at the lower-energy edge of the second band (point $M$ in Fig. \ref{dispersion}), and cannot be described within the framework of the discrete model.
\begin{figure}[hbt]
\includegraphics[width=8.0 cm]{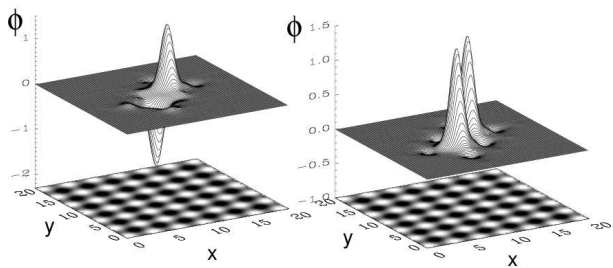}
\caption{\label{twisted} Spatial structure of the higher-order gap
solitons formed by two (left) out-of-phase, and (right)
in-phase fundamental modes ($V_0=5.0$, $\mu=4.0$). Filled contours show the lattice potential; dark shading corresponds to potential minima.}
\end{figure}
We have also found families of higher-order gap modes that can
be identified as bound states of the fundamental gap solitons. These states can be
centered on the lattice potential minima or maxima, similarly to
higher-order "odd" and "even" states in 1D lattices \cite{pearl}, and can exhibit
symmetry-breaking instabilities. Two examples of the lowest-order in- and out-of-phase odd modes,  centered on a lattice maximum, are shown in Fig. \ref{twisted}.

The crucial issue of the potential observation of 2D gap solitons
is the stability of the 2D localized state in a 3D BEC cloud, i.e.
dynamical stability of the 3D state
$\Psi(x,y,z;t)$ with the initial condition given by
$\Phi(z)\phi(x,y)$, where $\phi(x,y)$ is the two-dimensional
stationary gap soliton. The analysis of
stability of 2D solitons in a 3D BEC cloud, wich will provide clues for
possible experimental observation of multi-dimensional gap
solitons, is the subject of our separate study.

In conclusion, we have
demonstrated that the interaction of BEC with the lattice
potential is analogous to the light scattering by a nonlinear
photonic bandgap structure. We have studied the properties of two-dimensional atomic
bandgap structures and demonstrated the existence of
gap solitons, the spatially localized states of
BEC existing in the gaps of the matter-wave spectrum. We believe
the analogy between the physics of BEC in optical lattices and
photonic crystals can be useful for revealing many novel features of the
matter-wave dynamics in reconfigurable atomic bandgap structures. 

The authors thank C.M. Savage and J.J. Garc\'{\i}a-Ripoll for invaluable advice. This work is supported by the Australian Partnership for Advanced Computing and Australian Research Council. The authors are part of the ARC Centre for Quantum-Atom Optics.

\bibliography{2d_lattice}

\end{document}